 \documentclass[11pt,onecolumn]{IEEEtran}
    \usepackage{geometry}
\ifCLASSINFOpdf
\else
  \usepackage[dvips]{graphicx}
  \graphicspath{{../eps/}}
  \DeclareGraphicsExtensions{.eps}
\fi

\usepackage[cmex10]{amsmath}
\usepackage{mdwmath}
\usepackage{amssymb}  
\usepackage{textcomp}

\usepackage[tight,footnotesize]{subfigure}

\usepackage[font=footnotesize]{subfig}

\usepackage{stfloats}

\begin{document}

\title{Distributed Flow Scheduling in an Unknown Environment}

\author{
\IEEEauthorblockN{Yaoqing Yang \IEEEauthorrefmark{1}\IEEEauthorrefmark{2}, Keqin Liu \IEEEauthorrefmark{2}, Qing Zhao \IEEEauthorrefmark{2}}\\
 \IEEEauthorblockA{\IEEEauthorrefmark{1} Department of Electronic Engineering, Tsinghua University, Beijing, China}\\
 \IEEEauthorblockA{\IEEEauthorrefmark{2} Department of Electrical and Computer Engineering, UC Davis, California, USA}\\
 \IEEEauthorblockA{Email: yqyang1991@gmail.com}
}%

\markboth{Y. Yang, K. Liu, Q. Zhao}%
{Shell \MakeLowercase{\textit{et al.}}: Distributed Flow Scheduling in Unknown Environment}

\maketitle

\newcommand{\proofend}{\ensuremath{\rlap{\hbox{\rule[.4 ex]{0.2 ex}{0 ex}\rule[.4 ex]{0.84 ex}{0.07 ex}\rule[.4 ex]{0.07 ex}{0.9 ex}}}\square}}

\begin{abstract}
Flow scheduling tends to be one of the oldest and most stubborn problems in networking. It becomes more crucial in the next generation network, due to fast changing link states and tremendous cost to explore the global structure. In such situation, distributed algorithms often dominate. In this paper, we design a distributed virtual game to solve the flow scheduling problem and then generalize it to situations of unknown environment, where online learning schemes are utilized. In the virtual game, we use incentives to stimulate selfish users to reach a Nash Equilibrium Point which is valid based on the analysis of the `Price of Anarchy'. In the unknown-environment generalization, our ultimate goal is the minimization of cost in the long run. In order to achieve balance between exploration of routing cost and exploitation based on limited information, we model this problem based on Multi-armed Bandit Scenario and combined newly proposed DSEE with the virtual game design. Armed with these powerful tools, we find a totally distributed algorithm to ensure the logarithmic growing of regret with time, which is optimum in classic Multi-armed Bandit Problem. Theoretical proof and simulation results both affirm this claim. To our knowledge, this is the first research to combine multi-armed bandit with distributed flow scheduling.
\\
\\
\({\;\;\;\;}\)\textit{\textbf{Keywords--Flow Scheduling, Price of Anarchy, Multi-Armed Bandit, Logarithmic Regret}}
\\
\\
\end{abstract}

\section{\textbf{INTRODUCTION}}
\vspace{0.2cm}
We consider a network sharing optimization problem. All of the users would like to optimize their own path selection without exchanging information with others. However, congestion on the same edge introduces increasing cost. We would like to figure out a distributed scheme for them to find a best solution.

We assume here that each user has a flow with unit capacity requirement but different source or destination. However, generalization to multi-commodity situation is not difficult if we split flows into units and carry out the algorithm for each unit flow. Cost on each edge is a random variable due to link state changes and environment variances. As mentioned above, conflictions increase costs, so we assume the expectation of one such variable grows when flows routed on it increase. In the front half of this paper, we assume these expectations are known and we focus on the virtual game designing to find the flow scheduling scheme.

In the second half, we generalize our problem into unknown environment. That is, we do not know the expectations of edge costs and we need moderate exploration. We use the newly proposed DSEE Sequence[17] to optimize the time for exploration. After exploration, samples of edge costs are stored in routers and the sample means are calculated to approximate the expectations. Exploration periods happen periodically in a predetermined manner so routers know when to explore. Between two neighboring exploration periods is an exploitation period. At the beginning of an exploitation period, we use the distributed Bellman-Ford algorithm[16] to calculate routing tables based on the sample means. In order to solve the confliction problem, we apply the virtual game here. During the rest time of the exploitation period, we route flows according to the routing tables. Obviously, exploration and Bellman Ford periods both introduce extra cost, or reward loss. The ultimate object for us is to design a distributed algorithm to minimize long-run total cost for the whole network. In the whole paper, we assume that time is slotted and both explorations and exploitations need time.

\subsection{\textbf{Background of Flow Scheduling}}
\vspace{0.2cm}

Problems of flow scheduling in known scenario could still be very hard to solve. There are increasing literatures in this area with development of the widely-used MPLS network. Here, we base our work on background of flow scheduling instead of packet switching, wired or wireless, in order to make it more practical and useful nowadays.

The \textit{minimum interference routing} [1]-[4] is a prospective direction in flow scheduling. Its purpose can be quite similar with ours. However, minimum interference routing algorithms, like MIRA[2] and WSP[4], consider more about load balancing to maintain the sustainability of future flow admitting , while we want to solve an optimization problem right now. Extensions of our work approve of adaptive scheduling of newly admitted flow but all routers should be informed beforehand that new flows have come in.

Literatures in the Routing Games are more relevant to our problem. Firstly, our modeling is very similar to the modeling of the \textit{atomic routing} in [5]. Secondly, at the Bellman Ford period users perform a virtual game and take turns to select their own optimized routing path without considering congestion to others, which is the same with routing games. However, there is still fundamental difference between our virtual game and atomic routing. Firstly, we let distributed routers decide the best paths for the players, other than players select by themselves. This is more reasonable since in real life, routers decide paths for users. Secondly, our game is only virtual, which is used finally to solve an optimization problem. However, it is well known that games won't always converge to the optimum point. So we set the extra cost one user introduces to the whole network as the revenue he pays (see part II.B) to make this non-cooperative game a situation when selfish optimization equals social optimization. We prove the fast convergence to Nash Equilibrium Point in this routing game and use the constant bound of the `Price of Anarchy' to measure its worth[9]. Moreover, modeling of [5] does not consider the generalization to unknown environment, so our work is more general.

\subsection{\textbf{Stochastic online learning based on MAB Problem}}
\vspace{0.2cm}

Second half of our paper focuses on the generalization to unknown model. The nature of routing problem with unknown edge cost calls for introduction of the Multi-armed Bandit (MAB) Problem. In the classic MAB, there are N independent arms and one single player. Each arm, when played, incurs a random cost with an unknown distribution. The player should decide the sequence to play each arm to obtain the minimum cost. We notice that the player should try to maintain the balance between exploration and exploitation, which respectively means to play a new arm and learn its cost distribution and to play the arm with minimum cost. A frequently used criterion to judge the performance of an adopted sequence is the so called \textit{regret} or \textit{cost of learning}, defined as the difference in total cost between the chosen sequence and the optimum sequence when cost distribution is known. The best regret, logarithmically growing with time, is obtained in [10] by Lai and Robbins. In [11][12], authors gave out index-type policies to achieve logarithmic regret.

Routing problems with unknown edge cost distributions can be modeled as a variation of the classic MAB problem if we view each path as an arm. However, performances of classic algorithms degrade severely here since paths with shared edge cannot be viewed as independent. In [13], Liu and Zhao explore the dependence of paths to obtain a logarithmic regret. In [14], Gai and Krishnamachari made modifications to UCB1 [12] and applied their algorithm LLC into shortest path problem. However, none of them gave out distributed method for path selection. In our work, we put this difficulty into the design of a distributed virtual game and solve it beforehand in known model. It's important to note that the concept Distributed Learning in [15] is different from our concept of `distributed'.  `Distributed' in [15] means that each user does not exchange information with others and finds the best arm on his own. However, we further assume that our algorithm should be carried out distributedly in each router by using the Bellman Ford Algorithm. Moreover, [12]-[15] did not consider network sharing, so our work is more general.

In our paper, we explore an algorithm doing online learning for multi-user situation in a distributed way. To our knowledge, no previous work considered such comprehensive situation. Based on our algorithm, the whole network can also achieve logarithmic regret with time. However, in order to judge the virtual game at the same time, we define regret slightly differently from the classic definition.
\\
\\
\textbf{Definition 1: }We define \textbf{Regret} as the number of time slots when the network is not in a Nash Equilibrium Point.
\\
\\
In Regret Analysis part, we analyze the equivalence between definition 1 and the classic one. We prove that our virtual game reaches a Nash Equilibrium Point in limited circles, and regret grows logarithmically with time. These claims ensure the effectiveness of the virtual game.

It is important to note that the Optimum Point is also a Nash Equilibrium Point in our game. However, Nash Equilibrium Point is not unique since strategy domain for each user is discrete (different paths). Commonly, only when we have continuous strategy domain, Nash Equilibrium Point is unique[5][6]. So analysis of the \textit{Price of Anarchy} is necessary.

\section{\textbf{SYSTEM MODEL}}
\vspace{0.2cm}

\subsection{\textbf{Cost Modeling}}
\vspace{0.2cm}

Consider a graph \(G=(V,E)\) and \(K\) source-destination pairs \((s_k,t_k)\), each with unit amount \(f_k=1\). For each edge \(e\in E\), define flow on the edge
\begin{equation}
f_e=\sum_{e\in p_k}f_k
\end{equation}
 in which the \(p_k\) represents the path chosen by the \(k\)th flow. Since all flows have unit amount, the flow on each edge \(f_e\) will take discrete value from \(\{1,2,...,K\}\). Define
\begin{equation}
C(F)=\sum_{e\in E}c_{e}(f_{e})
\end{equation}
as the total cost in one time slot, in which the \(c_{e}\) represents the cost for edge $e$. At each time slot, for each edge $e$ and a certain flow amount \(f_{e}\), \(c_{e}(f_{e})\) is a random variable whose expectation value increases when \(f_{e}\) grows. For different time slots, \(c_{e}(f_{e})\) is an i.i.d. random process. \(F\) denotes the whole flow distribution on the network. In order to minimize the time average of \(C(F)\), we try to obtain the best flow distribution \(F\) in a distributed way to minimize the expectation of \(C(F)\). Henceforth we use a bar to represent the expectation. For example,  \(\bar{C}(F)\) denotes the expectation of \(C(F)\). The unit amount is the granularity of all flows. Obviously, generalization to multi-commodity scenario is trivial if we split flows into flow units and treat each unit as an independent flow.

\subsection{\textbf{Incentive}}
\vspace{0.2cm}

In the virtual game design, users are assumed selfish since they could not exchange information. In order to stimulate users to cooperate, we set revenues as incentives for them. Assume at some time \(t\), there are already \(K_t\) flows in the network and the whole flow distribution is currently \(F_t\). Then the whole cost of the network equals \(\bar{C}(F_t)\). For a certain \(kth\) flow, let \(F_t(k)\) denote the flow distribution when \(f_k\) is withdrawn from \(F_t\). Then we define
\begin{equation}
\bar{C}(F_t)-\bar{C}(F_t(k))
\end{equation}
as the revenue for the \(kth\) flow. We can easily see that when a user has the opportunity to change its routing path, he surely chooses the path that introduces the minimum extra cost to the whole network. Then the total cost decreases.

\section{\textbf{ALGORITHM IN KNOWN MODEL}}
\vspace{0.2cm}

\subsection{\textbf{Virtual Game Design}}
\vspace{0.2cm}
In this part, we assume that routers know all \(\bar{c}_e(f_e)\) beforehand. Each user takes turns to hire routers to do Bellman Ford Algorithm. The price for each edge is set as the incentive described in II.B.

There will be \(N*K\) time slots reserved for one circle. So time reserved for each user is N slots, and the Bellman Ford Algorithm surely converges in such long period. Also, total cost decreases each time when a user changes path, since the revenue for this user defined earlier is equal to the extra cost to the whole network introduced by him.
\\
\\The complete algorithm is as follows:

1) Take out the \(kth\) flow from current flow distribution. If it is the first time for this flow to do path optimization and routers do not know yet the path to transmit this flow, they do not need to take it out.

2) Calculate price on each edge. The price is the extra cost if this edge is chosen:
\begin{equation}
P_e(F)=\bar{c}_e(f_e)-\bar{c}_e(f_e-f_k)
\end{equation}

3) Start the Bellman Ford Algorithm and wait for N slots to ensure its convergence. The source node is \(s_k\). Find out the path with minimum price to transmit flow to \(d_k\).

4) Add up \(f_k\) on each edge chosen to transmit the \(kth\) flow.

5) Do the 1) again for the \(k+1 th\) flow.

\subsection{\textbf{Nash Equilibrium Point}}
\vspace{0.2cm}

\noindent \textbf{Theorem 1: }If we do algorithm described in III.A, then after finite circles, the whole network reaches a Nash Equilibrium Point. Convergence time is bounded.
\\
\\
\textbf{Proof: }During one circle, one of two events below must occur:

a).At least one user changes his path.

b).No one changes his path.

If event `b' happens, we know that no one could change his path unilaterally. Obviously the network has reached Nash Equilibrium Point.

However, if event `a' happens, total cost decreases. This has been stated in II.B. Since there will be limited paths for one flow to take, number of flow distribution is limited, too. So  `a' won't happen all the time.

We can further figure out the upper bound of convergence time to reach a Nash Equilibrium Point. In fact, we need \(\lceil\frac{S_M}{S_m}\rceil\) times of Bellman Ford circles. The \(S_M\) denotes the maximum difference between cost of two different flow distribution, and \(S_m\) denotes the minimum. This is true because during each Bellman Ford circle, the cost of the whole network will at least decrease by \(S_m\) if `b' does not happen. \proofend

\section{\textbf{PRICE OF ANARCHY}}
\vspace{0.2cm}

In this part we give out the analysis of the `Price of Anarchy'. This notion was originally defined in [8] to measure the selfish performance of a simple game of N players that compete for M parallel links. In [9], the authors analyzed the price of anarchy of an atomic routing game to polynomial edge cost with nonnegative coefficient. They gave out results of \(d^{O(d)}\) in which \(d\) represents the highest order of the polynomial edge cost function. This result is considered by [5] to be a significant generalization of previous work.

In our paper, we still need analysis of the `Price of Anarchy' since our ultimate goal is to solve an optimization problem. So far, we give out algorithm to make different users optimize their own price--the incentive--to reach a Nash Equilibrium Point. So we need to figure out the difference between a Nash Equilibrium Point and the optimum point.
\\
\\
\textbf{Definition 2: }We define the \textbf{Price of Anarchy} as
\begin{equation}
\bar{C}(F_N)/\bar{C}(F^{*})
\end{equation}
The \(F_N\) represents flow distribution of one Nash equilibrium point. And \(F*\) represents flow distribution of the optimum point, in which (2) is optimized.
\\
\\
We give out existence of constant price of anarchy for general polynomial edge cost. Then we give out concrete value for polynomials with nonnegative coefficients. Here we need the functions to be convex but this is trivial when congestion is concerned. The assumption of polynomial edge cost is common in previous work of Routing Games[5][8][9]. Our modeling is different from routing game. In Routing Games, the `total price' in (6) is equal to the expectation of total cost function defined in (2), while they are different in our virtual game. However, polynomial functions are quite enough to model congestions in our problem, so we still use this assumption.

\subsection{\textbf{General Polynomial Function: Existence}}
\vspace{0.2cm}

In this part we prove the existence of constant upper bound of the price of anarchy for polynomial edge cost function. In another word, this constant is independent of network size and topology. In the proof we use the following definition.
\\
\\
\textbf{Definition 3: }We define \textbf{Total Price} for distribution \(F\) as
 \begin{equation}
P(F)=\sum_{e\in E}[\bar{c}_{e}(f_{e})-\bar{c}_{e}(f_{e}-f_{u})]\cdot f_e
\end{equation}
\(f_u\) just means the unit flow amount.\\

We simply replace \(f_u\) with 1 in following parts, since we claim in section II.A that all flow has the same unit amount. What is important is the reason we define (6) as the `total price'. In fact, from (3)(4) we know that the incentive pricing scheme asks for the \(kth\) user a price of
 \begin{equation}
P_k(F)=\sum_{e\in p_k}[\bar{c}_{e}(f_{e})-\bar{c}_{e}(f_{e}-1)]
\end{equation}
We add up (7) for all users and we get
 \begin{equation}
P(F)=\sum_{k=1}^{K}\sum_{e\in p_k}[\bar{c}_{e}(f_{e})-\bar{c}_{e}(f_{e}-1)]
\end{equation}
Simply change the order of summation and we get (6).
\\
\\
\textbf{Theorem 2: }If the expectation of edge cost function \(\bar{c}_{e}(f_{e})\) is convex and grows polynomially with \(f_e\), there exists a constant bound for the `Price of Anarchy' independent of network size and flow amount.\\

We assume that edge cost functions are polynomials of maximum degree \(d\). Here \(d\) is different from the degree of barycentric spanner in proof of \textbf{Theorem 4}.
\begin{equation}
\bar{c}_e(f_e)=a_e f_e^d + \sum_{i=1}^{d} a_e^{(i)} f_e^{d-i}
\end{equation}

First we give out Lemma 1. This is the relationship between total cost and total price.
\\
\\
\textbf{Lemma 1: } For a given network G=(V,E), there exist two constant numbers \(A_l, A_r\). For any flow distribution \(F\), we have
\begin{equation}
A_l\le \frac{P(F)}{\bar{C}(F)}\le A_r
\end{equation}
These two numbers are independent of the network size.\\

The nature of Lemma 1 is very simple. For a polynomial \(E(c_e)\), the numerator and the denominator of (10) is of the same order of flow amount \(f_e\). So the fraction is certainly limited. We put detailed proof in Appendix A. Similarly, we could arrive at the following formula.\\

For a given G=(V,E), there exists a constant number \(A_u\). For any flow distribution and any edge \(e\), it satisfies
\begin{equation}
\frac{\bar{c}_e(f_e+1)-\bar{c}_e(f_e)}{\bar{c}_e(f_e)-\bar{c}_e(f_e-1)}\le A_u
\end{equation}\\

Then we give out \textbf{Lemma 2}. This is the `Variational Inequality Characterization'[5], which describes the basic feature of a Nash Equilibrium Point. Proof of \textbf{Lemma 2} is also put in the appendix.
\\
\\
\textbf{Lemma 2: } For a given network G=(V,E) and a Nash Equilibrium point \(F\) of K users, for any flow distribution \(F{'}\), we have
\begin{equation}
\begin{split}
\sum_{e\in E}[\bar{c}_e(f_e)-\bar{c}_e(f_e-1)]\cdot f_e \le  A_u\sum_{e\in E}[\bar{c}_e(f_e)-\bar{c}_e(f_e-1)]\cdot f_e{'}
\end{split}
\end{equation}\\

Based on these two Lemmas, we can complete the proof of \textbf{Theorem 2}. The proof is still very simple in nature. We have proven that the total cost(2) and the total price(6) grows with flow amount in the same order (\textbf{Lemma 1}). Then we find the constant upper bound of \(\frac{P(F_N)}{P(F*)}\) (\textbf{Lemma 2}). These two steps complete the proof. The detailed proof is put in Appendix C.

\subsection{\textbf{Polynomial Function with Nonnegative Coefficients: Concrete Value}}
\vspace{0.2cm}

For polynomial edge cost with nonnegative coefficients, we give out concrete value of the upper bound. Although we could derive a proof based on the same procedure of part IV.A, we can take advantage of the nonnegative coefficients to get a relatively simple proof in the Appendix. First we give out some definitions. If (8) holds and coefficients are all nonnegative, we have for each edge e
\begin{equation}
\begin{split}
\bar{c}_e(f_e+1)-\bar{c}_e(f_e)=a_e [(f_e+1)^d-f_e^d] +\sum_{i=1}^{d} a_e^{(i)} [(f_e+1)^{d-i}-f_e^{d-i}]
\end{split}
\end{equation}
Obviously, all terms in (13) have nonnegative coefficients. We assume
\begin{equation}
\begin{split}
\bar{c}_e(f_e+1)-\bar{c}_e(f_e)=\sum_{i=0}^d \tilde{a}_e^{(i)} f_e^{d-i-1}
\end{split}
\end{equation}
in which \(\tilde{a}_e^{(i)}>0\) and \(\tilde{a}_e^{(0)}=a_e\).
Moreover,
\begin{equation}
\begin{split}
\sum_{i=0}^d \tilde{a}_e^{(i)}=\bar{c}_e(1)-\bar{c}_e(0)
\end{split}
\end{equation}
We assume
\begin{equation}
s_e=\mathop{min}\limits_{a_e^{(i)}>0}(a_e^{(i)})
\end{equation}
\begin{equation}
L=\mathop{max}\limits_{e\in E}[\bar{c}_e(1)-\bar{c}_e(0)]
\end{equation}
\\
\textbf{Theorem 3} For a given network G=(V,E), if all edge cost functions satisfy (9) and coefficients are nonnegative, we could give out the concrete value of the constant upper bound of the Price of Anarchy. The constant is \([(d+1)L\mathop{max}\limits_{e\in E}\frac{1}{s_e}]^d=d^{O(d)}\).

\section{\textbf{ALGORITHM IN UNKNOWN MODEL}}
\vspace{0.2cm}

From this section, we give out generalization to unknown model. In another word, we further assume that the cost distribution of each edge is unknown at the beginning. In order to get enough information about the network, we adopt the newly proposed DSEE Sequence algorithm in [17] and cut time into interleaving exploration and exploitation periods. A router sends exploration flows to get samples of the cost and store them in memory. Based on these samples, a router calculates sample mean and view it as the expectation of edge cost when doing Bellman-Ford Algorithm. Between the exploration periods are the exploitation periods, at the beginning of which the virtual game is applied. During the rest time of exploitation, users share the network based on routing tables. In order not to route flows on edges with high price, each user consents to do enough explorations. However, exploration periods and Bellman Ford periods cannot be too long since they introduce extra cost to the network.

\subsection{\textbf{Exploration}}
\vspace{0.2cm}

One exploitation period lasts for \(N=|V|\) time slots. In one exploration period, only one source node \(s_k\) starts exploration. K source nodes take turns to do exploration in different exploration periods. At the beginning of the first exploration period, \(s_1\) sends out a short flow of a random amount \(k_1\) to a random edge \(e_r\) related to it to explore the value \(c_{e_r}(k_1)\). Then the other node of edge \(e_r\) receives this flow and forward it in the next time slot. This whole exploration period terminate in \(N=|V|\) time slots. In the next exploration period, the source node \(s_2\) starts exploration instead of \(s_1\). The constant number \(N=|V|\) is large enough to ensure a minimum probability \(r=min_{e\in E}(r_e)>0\), in which the \(r_e\) is the probability of the edge \(e\) being estimated.

\subsection{\textbf{Exploitation}}
\vspace{0.2cm}

At the beginning of this period, there will be \(N*K\) time slots reserved for a Bellman Ford period. During one period, we do one circle of the virtual game described in III.A.

However, we should replace (4) with
\begin{equation}
P_e(F)=\hat{c_e}(f_e)-\hat{c_e}(f_e-f_k)
\end{equation}
in which \(\hat{c_e}(f_e)\) denotes the sample means stored in routers' memory.

\subsection{\textbf{DSEE}}
\vspace{0.2cm}

Time is divided into interleaving sequence of Exploration and Exploitation. At the beginning of each exploitation period, there is  \(N*K\) time slots arranged for Bellman Ford period to do virtual game. One Bellman Ford period terminates only when the total time \(N*K\) is reached. Similarly, one Exploration period ends after \(N\) time slots. However, the exploitation period ends when the time slot t satisfies
\begin{equation}
card(t)<Glog(t)
\end{equation}
in which the \(card(t)\) represents number of time slots used to do exploration up to time t. Certainly, the whole DSEE Sequence is determined beforehand once the parameter G has been chosen.

\section{\textbf{REGRET ANALYSIS}}
\vspace{0.2cm}

We define regret as the number of time slots when all the flows are not routed in Nash Equilibrium Point (see the end of the Introduction part). In section III.B, we have proved the inevitability for K users to reach the Nash Equilibrium Point in limited circles of virtual game. In this part, we analyze the equivalence of definition 1 with classic one. Then we prove regret grows logarithmically with time.

\subsection{\textbf{Equivalence between Definition 1 and classic definition}}
\vspace{0.2cm}

Classic definition of regret is the difference in total cost between the chosen strategy sequence and the optimum strategy sequence when cost distribution is known.

In our algorithm, there exist two conditions that regret increases. The first one is exploration or Bellman Ford. During these periods, no flows are transmitted. However, if we define an extra constant cost for each of such slot to get a classic definition, we can see that this two regrets grow with time in the same order. The second one is when flows are not routed in a Nash Equilibrium Point in an exploitation period. But in one such slot, extra cost cannot be larger than \(S_M\). Therefore, even if we define a classic regret, it still grows with same order of time.

The only difference is the distance from one Nash Equilibrium Point to the Optimum Point. However, finding the Optimum Point for different flows tends to be NP hard and it cannot be done in a distributed way. So we choose to define regret based on a sub-optimal Nash Equilibrium Point which cannot be further improved in a distributed manner. Previous parts have shown the constant `Price of Anarchy' bound, which convince of the feasibility of our definition.

\subsection{\textbf{Regret Order}}
\vspace{0.2cm}

\textbf{Theorem 4: }If the chosen G in (19) satisfies
\begin{equation}
G\ge max(3/r,\frac{8d^2|E|\sigma^2}{rc^2})
\end{equation}
then regret(T) increases with the form \(O(log(T))\).\\

Here we give out some definitions in Theorem 4.
\\
\\
\textbf{Definition 4: }Let \(S\) be a d-dimensional vector space. A set \(B=\{x_1,x_2,...,x_d\}\subset S\) is called a barycentric spanner for \(S\) if every \(x\) in \(S\) can be written as linear combination of elements of \(B\) with coefficients in \([-1,1]\).\\

It is shown in [15] that if \(S\) is a compact set, then it has a barycentric spanner. We know that the set of different paths for a certain source-destination pair \((s_k,d_k)\) is a compact vector space, thus it has a barycentric spanner with dimension \(d_k\). We assume \(d=\mathop{max}\limits_{k=1\sim K} d_k\). \(\sigma^2\) is the largest variance of all the edge cost under different flow distributions. \(r\) is the minimum of the probability that a certain edge is chosen during explorations. \(c_k\) is the minimum price difference between two paths for the \(kth\) user under all different flow distributions. Since number of flow distributions is limited, \(c_k\) surely exists. Then we can define\(c=\mathop{min}\limits_{k=1\sim K} c_k\). These parameters are all related to the network topology and can be obtained beforehand. However, while choosing a G based on (20) is doable, usually we can choose a smaller G. Here we only concern about the existence of a sufficient condition.

Proof of \textbf{Theorem 4} still can be found in the Appendix. Instead we give out the basic idea of the proof. If G is chosen big enough, sufficient times will be used for exploration so that we have relatively accurate sample means for the cost of each edge under different flow amount. Based on Bernstein's inequality, we can bound the variance of sample means of path cost. When this variance is small enough, we can bound the probability that we make mistakes in the virtual game circle. Mistake-free virtual game results in Nash Equilibrium. Although proof of \textbf{Theorem 4} seems lengthy, it relies on this simple idea.

\section{\textbf{SIMULATIONS}}
\vspace{0.2cm}

\subsection{\textbf{Price of Anarchy Simulation}}
\vspace{0.2cm}
In this part we give out simulation result for the `Price of Anarchy'. Figure 1 shows the probability density function of the `Price of Anarchy' for different cost function orders. Large density near price 1 proves the efficiency of our algorithm. Also, the relationship between the `Price of Anarchy' and cost function order can be observed: distribution with a higher order has a longer tail.

\begin{figure}[hp]
\centering
\includegraphics[scale=0.8]{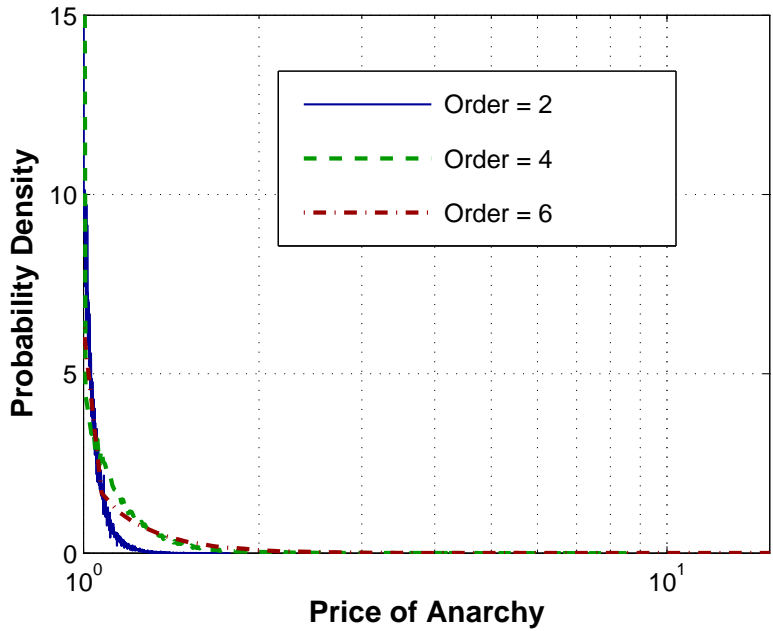}
\caption{`Price of Anarchy' distribution}\label{fig:graph}
\end{figure}

\subsection{\textbf{Regret Simulation}}
\vspace{0.2cm}
In this part we give out the simulation results for regret order. Figure 2 shows the growing behavior of regret with time under different G selections. We choose the \(G_b\) as the basic G based on the condition shown in \textbf{Theorem 4}. Actually, this condition is just an sufficient condition that leads to logarithmic growing of regret. In real simulation, we have chosen a basic G smaller than in \textbf{Theorem 4} but can still help the logarithmic growth hold.

\begin{figure}[hp]
\centering
\includegraphics[scale=0.8]{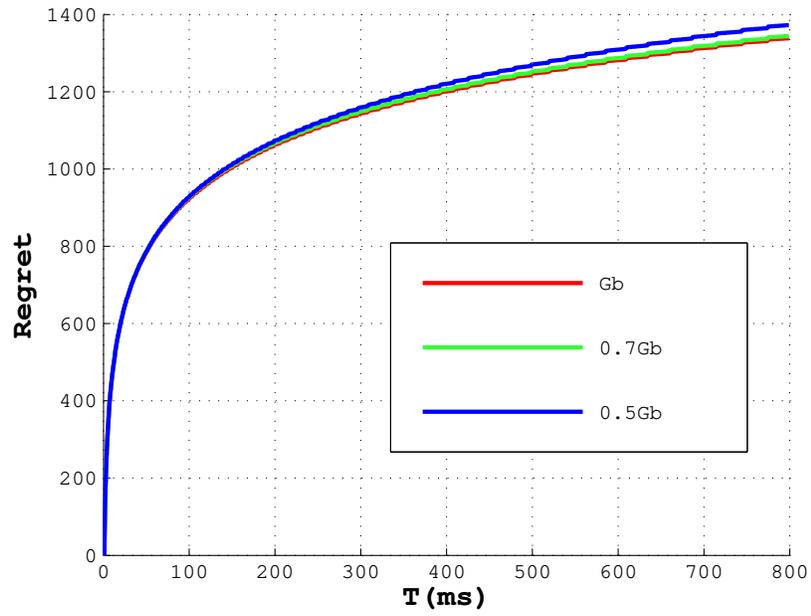}
\caption{regret(T)}\label{fig:graph}
\end{figure}

The second figure is the regret divided by log(T). It could help us see more clearly how the regret converges to a logarithmic order. Moreover, we see from simulation that if G is not large enough, the regret grows with an order larger than log(T). So in real-life applications, we should make sure that G is large enough.

\begin{figure}[hp]
\centering
\includegraphics[scale=0.8]{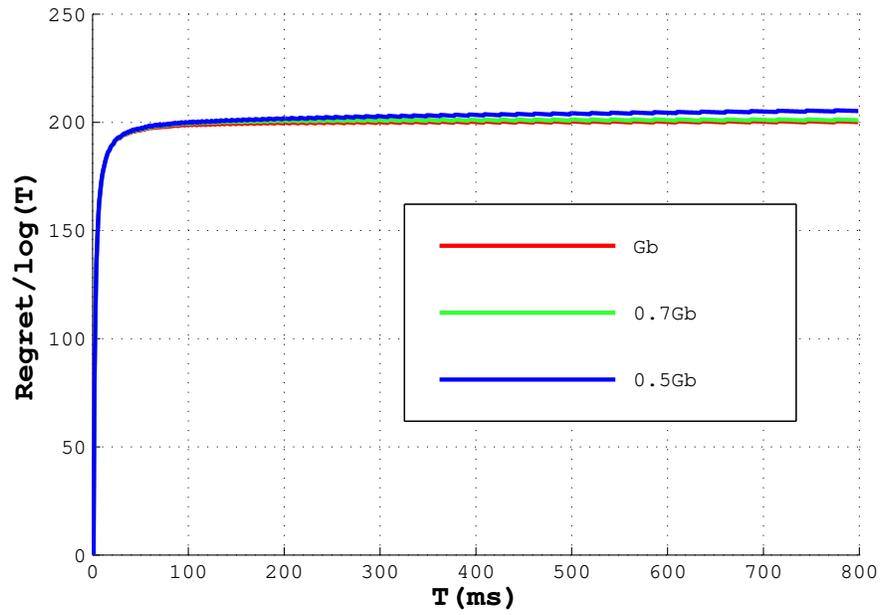}
\caption{regret(T) divided by log(T)}\label{fig:graph}
\end{figure}

\section{\textbf{Conclusions}}

In this paper, we considered the flow scheduling problem both under known and unknown model. For the known model, we proposed a virtual non-cooperative game with incentive pricing to solve cost optimization problem for users who do not exchange information with each other. To analyze this virtual game, we proved the fast convergence of the game into a Nash Equilibrium Point which had a bounded price of anarchy. The constant bound was proved to be independent of network size and flow amount. Then we extended this algorithm to situations when cost distributions were unknown beforehand. We modeled this problem under multi-armed bandit model and combined the virtual game with the newly proposed DSEE Sequence which could achieve best regret for all light-tail cost distributions. Sure enough, regret of our algorithm was proved to be growing logarithmically with time if the DSEE parameters were chosen properly, which is best in the classic online learning scenario. Also, simulation results of the `Price of Anarchy' and the regret growing behavior were given out to test the essential correctness of all our claims.

\appendices
\section{Proof of \textbf{Lemma 1}}
Based on (9) we have

\begin{equation}
[\bar{c}_e(f_e)-\bar{c}_e(f_e-1)]\cdot f_e=a_e f_e^{d} + \sum_{i=1}^{d} b_e^{(i)} f_e^{d-i}
\end{equation}

Here \(a_e^{(i)}\) and \(b_e^{(i)}\) are coefficients. We do not require them to be nonnegative here, but in \textbf{Theorem 3}, we require \(a_e^{(i)}\) to be nonnegative. Divide (21) with \(f_e^d\) and we get
\begin{equation}
\frac {\bar{c}_e(f_e)-\bar{c}_e(f_e-1)}{f_e^{d-1}}=a_e + \sum_{i=1}^{d} b_e^{(i)} f_e^{-i}
\end{equation}

For any \(\epsilon > 0\), there exists a \(f_{e,\epsilon}\). For any \(f_e > f_{e,\epsilon}\), \begin{equation}
|\frac {\bar{c}_e(f_e)-\bar{c}_e(f_e-1)}{f_e^{d-1}}-a_e|<\epsilon
\end{equation}
So we have, for any \(f_e > f_{e,\epsilon}\),
\begin{equation}
a_e-\epsilon< \frac {\bar{c}_e(f_e)-\bar{c}_e(f_e-1)}{f_e^{d-1}}<a_e+\epsilon
\end{equation}
Since \(f_{e,\epsilon}\) is limited, there exists a closed section \(I_e\). For any \(f_e \le f_{e,\epsilon}\),
\begin{equation}
\frac {\bar{c}_e(f_e)-\bar{c}_e(f_e-1)}{f_e^{d-1}} \in I_e
\end{equation}
Since \(\frac {\bar{c}_e(f_e)-\bar{c}_e(f_e-1)}{f_e^{d-1}}>0\), \(0 \notin I_e \). We choose \(\epsilon < \frac{a_e}{2}\), and denote \(J_e = I_e \cup [\frac{a_e}{2},\frac{3a_e}{2}]\) and we have for any \(f_e\),
\begin{equation}
\frac {\bar{c}_e(f_e)-\bar{c}_e(f_e-1)}{f_e^{d-1}} \in J_e
\end{equation}

Similarly, we divide (9) with \(f_e^{d}\) and finally get

\begin{equation}
\frac {\bar{c}_e(f_e)}{f_e^{d}} \in J_e^{'}
\end{equation}

Here \(J_e\) and \(J_e^{'}\) are both closed sections excluding zero.
Then for any flow distribution \(F\), we have

\begin{equation}
\begin{split}
\frac{P(F)}{\bar{C}(F)}=&\frac{\sum_{e \in E} [\bar{c}_e(f_e)-\bar{c}_e(f_e-1)]\cdot f_e}{\sum_{e \in E} \bar{c}_e(f_e)}
\\=&\frac{\sum_{e \in E} [\bar{c}_e(f_e)-\bar{c}_e(f_e-1)]/ f_e^{d-1}}{\sum_{e \in E} \bar{c}_e(f_e)/ f_e^d}
\end{split}
\end{equation}

From (26)(27) we know there exist two numbers \(A_l, A_r\), for any flow distribution, (10) holds.
\proofend

\section{Proof of \textbf{Lemma 2}}
For a certain \(k\in \{1,2,...,K\}\), the Nash Equilibrium Point \(F\) satisfies
\begin{equation}
\sum_{e\in {p_k^N}}[\bar{c}_e(f_e)-\bar{c}_e(f_e-1)]\le \sum_{p_k \in \Gamma_k} \sum_{e\in p_k} [\bar{c}_e(f_e+1)-\bar{c}_e(f_e)] \cdot f_{p_k}
\end{equation}
Here \(\Gamma_k\) represents the set of all paths available to the \(kth\) user. And \(f_{p_k}=1\) only when the path \(p_k \in \Gamma_k\) is chosen by the \(kth\) user. Otherwise it equals zero. Obviously, (29) can be derived directly from the definition of Nash Equilibrium Point. For different path selection schemes, \(f_{p_k}\) varies. However, (29) always holds. For a certain flow distribution F', we add up (29) for all K users and get
\begin{equation}
P(F_N)\le\sum_{e\in E}[\bar{c}_e(f_e+1)-\bar{c}_e(f_e)] \cdot f_e{'}
\end{equation}
Since we have (11) already, we can get (12).\proofend

\section{Proof of \textbf{Theorem 2}}
Let F represents a random Nash Equilibrium point and \(F^{*}\) denotes the optimum point. For a certain edge e, from (26), we have
\begin{equation}
\frac{[\bar{c}_e(f_e)-\bar{c}_e(f_e-1)]/f_e^{d-1}}{[\bar{c}_e(f_e^{*})-\bar{c}_e(f_e^{*}-1)]/(f_e^{*})^{d-1}} \le \frac{J_e^{(L)}}{J_e^{(R)}}
\end{equation}
The \(J_e^{(L)}\) and \(J_e^{(R)}\) are left and right border of \(J_e\). And * represents the optimum point. From this inequality we can derive directly and get
\begin{equation}
\begin{split}
[\bar{c}_e(f_e)-\bar{c}_e(f_e-1)] \cdot f_e^{*}\le & (\frac{J_e^{(L)}}{J_e^{(R)}})^{\frac{1}{d}}\{[\bar{c}_e(f_e)-\bar{c}_e(f_e-1)] \cdot f_e\}^{\frac{d-1}{d}}
\\ &\cdot \{[\bar{c}_e(f_e^{*})-\bar{c}_e(f_e^{*}-1)] \cdot f_e^{*}\}^{\frac{1}{d}}
\end{split}
\end{equation}
Based on the \(H\ddot o\ lder\) inequality, we get
\begin{equation}
\begin{split}
\sum_{e \in E} [\bar{c}_e(f_e)-\bar{c}_e(f_e-1)] \cdot f_e^{*}
 \le &(\frac{J_e^{(L)}}{J_e^{(R)}})^{\frac{1}{d}}\sum_{e \in E} \{[\bar{c}_e(f_e)-\bar{c}_e(f_e-1)] \cdot f_e\}^{\frac{d-1}{d}}
\\ &\cdot \{[\bar{c}_e(f_e^{*})-\bar{c}_e(f_e^{*}-1)] \cdot f_e^{*}\}^{\frac{1}{d}}
\\ \le &(\frac{J_e^{(L)}}{J_e^{(R)}})^{\frac{1}{d}}\{\sum_{e \in E}[\bar{c}_e(f_e)-\bar{c}_e(f_e-1)] \cdot f_e\}^{\frac{d-1}{d}}
\\ &\cdot \{\sum_{e \in E}[\bar{c}_e(f_e^{*})-\bar{c}_e(f_e^{*}-1)] \cdot f_e^{*}\}^{\frac{1}{d}}
\\=& (\frac{J_e^{(L)}}{J_e^{(R)}})^{\frac{1}{d}} \cdot [P(F)]^{\frac{d-1}{d}} \cdot [P(F^{*})]^{\frac{1}{d}}
\end{split}
\end{equation}

Since (12) holds for every flow distribution \(F{'}\), we could let \(F{'}=F^{*}\) so
\begin{equation}
P(F) \le A_u \cdot (\frac{J_e^{(L)}}{J_e^{(R)}})^{\frac{1}{d}} \cdot [P(F)]^{\frac{d-1}{d}} \cdot [P(F^{*})]^{\frac{1}{d}}
\end{equation}
It means
\begin{equation}
P(F)/P(F^{*}) \le (A_u)^d \cdot \frac{J_e^{(L)}}{J_e^{(R)}}
\end{equation}

And we have Lemma 1, so we finally get
\begin{equation}
\begin{split}
\bar{C}(F)/\bar{C}(F^{*})= &\frac{\bar{C}(F)}{P(F)} \cdot \frac{P(F)}{P(F^{*})} \cdot \frac{P(F^{*})}{\bar{C}(F^{*})} \\ \le &(A_u)^d \cdot \frac{A_r}{A_l} \frac{J_e^{(L)}}{J_e^{(R)}}
\end{split}
\end{equation}

From previous Lemmas, we know absolutely that constants on the right side of this inequality are independent from network topology and flow distribution. Since \(F_N\) is a random flow distribution, we have proved \textbf{Theorem 2}. \proofend

\section{Proof of \textbf{Theorem 3}}
Conditions in this theorem also ensure the functions are convex. So we have for any flow distribution
\begin{equation}
\bar{C}(F)\le P(F)
\end{equation}

From (15) we have, for any i and any \(e\in E\)
\begin{equation}
\tilde{a}_e^{(i)}<L
\end{equation}
Based on \(H\ddot{o}lder\) inequality, we have
\begin{equation}
\begin{split}
\sum_{e \in E}[\bar{c}_e(f_e+1)-\bar{c}_e(f_e)]\cdot f_e^{*}=&\sum_{i=0}^d\sum_{e \in E} \tilde{a}_e^{(i)} f_e^{d-i-1}f_e^{*}
\\\le&\sum_{i=0}^d\{\sum_{e \in E} \tilde{a}_e^{(i)}( f_e^{d-i-1})^{\frac{d-i}{d-i-1}}\}^{\frac{d-i-1}{d-i}}
\\& \;\;\;\;\cdot\{\sum_{e \in E} \tilde{a}_e^{(i)}(f_e^{*})^{d-i}\}^{\frac{1}{d-i}}
\\\le& L\sum_{i=0}^d\{\sum_{e \in E} \frac{1}{s_e} \bar{c}_e(f_e) \}^{\frac{d-i-1}{d-i}}\{\sum_{e \in E} \frac{1}{s_e} \bar{c}_e(f_e^{*}) \}^{\frac{1}{d-i}}
\\\le& L\mathop{max}\limits_{e\in E}\frac{1}{s_e}\cdot \sum_{i=0}^d\{ \bar{C}(F) \}^{\frac{d-i-1}{d-i}}\{\bar{C}(F^{*})\}^{\frac{1}{d-i}}
\end{split}
\end{equation}
Since \(\bar{C}(F^{*})\le \bar{C}(F)\), we have
\begin{equation}
\begin{split}
\sum_{e \in E}[\bar{c}_e(f_e+1)-\bar{c}_e(f_e)]\cdot f_e^{*}\le (d+1)L\mathop{max}\limits_{e\in E}\frac{1}{s_e}\cdot \{ \bar{C}(F) \}^{\frac{d-1}{d}}\{\bar{C}(F^{*})\}^{\frac{1}{d}}
\end{split}
\end{equation}
For one random Nash equilibrium \(F\) and the optimum point \(F^{*}\), from (30)(37) we have
\begin{equation}
\bar{C}(F)\le P(F)\le\sum_{e\in E}[\bar{c}_e(f_e+1)-\bar{c}_e(f_e)] \cdot f_e^{*}
\end{equation}
Combining (16)(17)(40)(41) we have
\begin{equation}
\bar{C}(F) / \bar{C}(F^{*}) \le [(d+1)L\mathop{max}\limits_{e\in E}\frac{1}{s_e}]^d=d^{O(d)}
\end{equation}
And this constant is independent of network topology and flow distribution.\proofend

\section{Proof of \textbf{Theorem 4}}

Since the number of time slots used in exploration and Bellman Ford increases strictly with \(O(logt)\), we could only focus on the number of slots that all flows are not operating at the Nash equilibrium point. Define the \(A_t\) the event that all the flows are not operating at the Nash equilibrium point at time t. We give out the upper bound of \(P(A_t)\).
\\
\\
Define \(B_t^k\) as the event that  last Bellman Ford just before time slot t for the \(k\)th flow goes wrong since poor estimation of the path cost. Then
\begin{equation}
P(B_t^k)=P\{\hat{X}^*(t)\ge min_{p\in P}\hat{X_p}(t)\}
\end{equation}

The \(P\) denotes the set of paths that the \(k\)th flow can choose from. The \(\hat{X_p}(t)\) is the incentive price for choosing path \(p\). This price is calculated by adding up all the extra edge cost introduced by the \(k\)th flow. That is
\begin{equation}
\hat{X_p}(t)=\sum _{e\in E}\hat{c_e}(f_e)-\hat{c_e}(f_e-f_k)
\end{equation}
The \(p^*\) represents the real best path for \(k\)th flow to choose if price expectation for each edge is known exactly. And the \(\hat{X}^*(t)\) is the estimated price for choosing this path.
\\
\\
Let \(n_e(k,t)\) be the number of times \(e\in E\) is observed when the \(k\) units of flow are put on it up to time t during the exploration slots. Let \(r_e(k)\) represents the probability that \(e\) with flow \(k\) on it is chosen to be observed at a random time slot. Since \(k\) can only take values from \(\{1,2,...,K\}\) and the number of edges is limited, we can ensure the existence of \(r=\mathop{min}\limits_{e\in E}r_e\).
\\
\\
Obviously,
\begin{equation}
E(n_e(k,t))=Gr_e(k)logt
\end{equation}
\begin{equation}
Var(n_e(k,t))<Gr_e(k)logt
\end{equation}

so, based on Bernstein's inequality
\begin{equation}
\begin{split}
P\{n_e(k,t)<{\frac {1}{2}}Grlogt\}\le &P\{n_e(k,t)<{\frac {1}{2}}Gr_e(k)logt\}
\\<&exp(-\frac{1}{2}\frac{E^2(n_e(k,t))}{\frac{1}{2}E(n_e(k,t))+Var(n_e(k,t))})
\\=&t^{-\frac{1}{3}Gr_e(k)}\le t^{-1}
\end{split}
\end{equation}

Let \(M=\frac {1}{2}Grlogt\) and we can easily get
\begin{equation}
\begin{split}
P\{ \exists e \in E, k \in \{1,2,...,K\}, s.t.n_e(k,t)<M \}<\sum_{e\in E,1\le k \le K} P\{n_e(k,t)<M\} < K|E|t^{-1}
\end{split}
\end{equation}

We choose a barycentric spanner in the network and assume it has \(d_k\) elements \(\{p_1,p_2,...,p_{d_k}\}\), then
\begin{equation}
\begin{split}
\{\hat{X}^*(t)\ge min_{p\in P}\hat{X_p}(t)\}\subseteq&\{\hat{X}^*(t)-X^*(t)>\frac{c}{2}\}\cup_{l=1}^{d_k}\{\hat{X_l}(t)-X_l(t)<-\frac{c}{2d_k}\}
\end{split}
\end{equation}
in which
\begin{equation}
X_l(t)=\sum_{e\in p_l}[\bar{c}_e(f_e+f_k)-\bar{c}_e(f_e)]
\end{equation}
and \(X^*(t)\) represents the real minimum expectation price of the path for \(k\)th flow.
\\
\\
Specifically for each \(p_l\) we have
\begin{equation}
\begin{split}
\;\;\hat{X_l}(t)-X_l(t)=\sum_{e\in p_l}[\hat{c_e}(f_e+f_k)-\hat{c_e}(f_e)]-\sum_{e\in p_l}[\bar{c}_e(f_e+f_k)-\bar{c}_e(f_e)]
\end{split}
\end{equation}
When enough times are used to estimate each edge, the value above will have a high probability to be small. Let \(L_l\) denote the number of edges in \(p_l\). Then we have
\\
\begin{equation}
\begin{split}
&{\;\;}P\{\hat{X_l}(t)-X_l(t)<-\frac{c}{2d_k} | \forall e \in E, k \in \{1,2,...,K\}, n_e(k,t) \ge M \}
\\<&P\{|\hat{X_l}(t)-X_l(t)|>\frac{c}{2d_k} | \forall e \in E, k \in \{1,2,...,K\}, n_e(k,t) \ge M \}
\\<&\;\;\;\;\sum_{e \in p_l}P\{|\hat{c_e}(f_e)-\bar{c}_e(f_e)|>\frac{c}{2d_k L_l}
\\&\;\;\;\;\;\;\;\;\;\;\;\;\;\; |\forall e \in E, k \in \{1,2,...,K\}, n_e(k,t) \ge M \}
\\&+\sum_{e \in p_l}P\{|\hat{c_e}(f_e+f_k)-\bar{c}_e(f_e+f_k)|>\frac{c}{2d_k L_l}
\\&\;\;\;\;\;\;\;\;\;\;\;\;\;\; |\forall e \in E, k \in \{1,2,...,K\}, n_e(k,t) \ge M \}
\\\le& 2L_l*2exp(-\frac{1}{2}\frac{(\frac{c}{2d_k L_l})^2}{\frac{\sigma ^2}{Grlogt}})
\\\le &4|E|exp(-\frac{1}{2}\frac{(\frac{c}{2d|E|})^2}{\frac{\sigma ^2}{Grlogt}})
\\\le &4|E|t^{-1}
\end{split}
\end{equation}
\\
Similar upper bound of \(\hat{X}^*(t)\) can also be obtained. After that we get
\begin{equation}
P(B_t^k)<4(|E|+|E|^2)t^{-1}+t^{-1}<5|E|^2t^{-1}
\end{equation}
\\
Each event \(B_t^k\) leads to the event \(A_{\widetilde{t}}\) for some \(\widetilde{t}>t\). If we would like to make the whole K flows reach the Nash Equilibrium point, we should ensure that \(B\) does not happen for a period long enough before time \(t\). In fact, if \(B\) does not happen, we will need \(\lceil\frac{S_M}{S_m}\rceil\) circles of Bellman Ford period to do virtual game. This result is based on Theorem 1. This is because if  \(B\) does not happen, it tends to be the same situation that routers know exactly the cost distribution of each edge.
\\
\\
The nature of DSEE Sequence makes the start point of each exploration period in an exponential sequence. We present this fact in a heuristic way. For the start time \(t_1\) of a exploration period, we have
\begin{equation}
card(t_1)=Glogt_1
\end{equation}
and for the start point \(t_2\) of  the next exploration period we have
\begin{equation}
card(t_2)=Glogt_2
\end{equation}
Since \(card(t_1)+NK=card(t_2)\), we have
\begin{equation}
\frac{t_2}{t_1}=exp(\frac{NK}{G})
\end{equation}
\\Let \(\{t_1,t_2,...t_{\lceil\frac{S_M}{S_m}\rceil}\}\) denote the starting points of last \(\lceil\frac{S_M}{S_m}\rceil\) circles of Bellman Ford period before time \(t\). And let \(t_{\lceil\frac{S_M}{S_m}\rceil+1}\) denote the starting point of the following period after time \(t\).  We see obviously that
\begin{equation}
\frac{t_{\lceil\frac{S_M}{S_m}\rceil+1}}{t_1}=exp(\frac{NK{\lceil\frac{S_M}{S_m}+1\rceil}}{G})
\end{equation}
For any Bellman Ford time slot \(t^*\) between \(t_1\) and \(t_{\lceil\frac{S_M}{S_m}\rceil+1}\) ,it satisfies that
\begin{equation}
\begin{split}
\frac{t}{t^*}<\frac{t_{\lceil\frac{S_M}{S_m}\rceil+1}}{t_1}=exp(\frac{NK{\lceil\frac{S_M}{S_m}+1\rceil}}{G})
\end{split}
\end{equation}
During these circles of Bellman Ford period, if B does not happen, the \(A_t\) does not happen either.
So we have
\begin{equation}
\begin{split}
P(A_t)<&\sum_{t^*,k=1,2,...,K}P(B_{t^*}^k)<\sum_{t^*,k=1,2,...,K}5|E|^2{(t^*)}^{-1}
\\<&10K|E|^2 \lceil \frac{S_M}{S_m} \rceil exp(\frac{NK\lceil\frac{S_M}{S_m}+1\rceil}{G}){t}^{-1}
\end{split}
\end{equation}
\\
In another word, the total regret to time horizon T can be written as
\begin{equation}
\sum_{t=1}^T{P(A_t)}=\sum_{t=1}^T{O(t^{-1})}
\end{equation}
and it is \(O(logT)\)\proofend

\ifCLASSOPTIONcaptionsoff
  \newpage
\fi


\end{document}